\documentclass[aps,prl,superscriptaddress,twocolumn]{revtex4-1}

\usepackage{graphics}
\usepackage{graphicx} 
\usepackage{dcolumn}
\usepackage{bm}
\usepackage{epstopdf}

\usepackage{amssymb,amsmath}
\usepackage{color}
\usepackage{array}

\newcommand{\jeff}{${J}_{\mathrm{eff}}\mathbf{=}1/2$~}
\newcommand{\jeffthree}{${J}_{\mathrm{eff}}\mathbf{=}3/2$~}

\begin{document} 




\title{Strongly Gapped Spin-Wave Excitation in the Insulating Phase of NaOsO$_3$}



\author{S. Calder}
\email{caldersa@ornl.gov}
\affiliation{Quantum Condensed Matter Division, Oak Ridge National Laboratory, Oak Ridge, Tennessee 37831, USA}

\author{J. G. Vale}
\email{j.vale@ucl.ac.uk}
\affiliation{London Centre for Nanotechnology and Department of Physics and Astronomy, University College London, Gower Street, London, WC1E 6BT, United Kingdom}
\affiliation{Laboratory for Quantum Magnetism, Ecole Polytechnique F\'ed\'erale de Lausanne (EPFL), CH-1015, Switzerland}

\author{N. Bogdanov}
\affiliation{Institute for Theoretical Solid State Physics, IFW Dresden, D01171 Dresden, Germany}

\author{C. Donnerer}
\affiliation{London Centre for Nanotechnology and Department of Physics and Astronomy, University College London, Gower Street, London, WC1E 6BT, United Kingdom}

\author{D. Pincini}
\affiliation{London Centre for Nanotechnology and Department of Physics and Astronomy, University College London, Gower Street, London, WC1E 6BT, United Kingdom}
\affiliation{Diamond Light Source Ltd, Diamond House, Harwell Science and Innovation Campus, Didcot, Oxfordshire OX11 0DE, United Kingdom}

\author{M. Moretti Sala}
\affiliation{European Synchrotron Radiation Facility, BP 220, F-38043 Grenoble Cedex, France}

\author{X. Liu}
\affiliation{Condensed Matter Physics and Materials Science Department and National Synchrotron Light Source-II, Brookhaven National Laboratory, Upton New York 11973, USA}
\affiliation{Beijing National Laboratory for Condensed Matter Physics and Institute of Physics, Chinese Academy of Sciences, Beijing 100190, China}

\author{M. H. Upton}
\affiliation{Advanced Photon Source, Argonne National Laboratory, Argonne, Illinois 60439, USA}

\author{D. Casa}
\affiliation{Advanced Photon Source, Argonne National Laboratory, Argonne, Illinois 60439, USA}

\author{Y.~G.~Shi}
\affiliation{Beijing National Laboratory for Condensed Matter Physics and Institute of Physics, Chinese Academy of Sciences, Beijing 100190, China} 
\affiliation{Research Center for Functional Materials, National Institute for Materials Science, 1-1 Namiki, Tsukuba, Ibaraki 305-0044, Japan}

\author{Y. Tsujimoto}
\affiliation{Research Center for Functional Materials, National Institute for Materials Science, 1-1 Namiki, Tsukuba, Ibaraki 305-0044, Japan}

\author{K.~Yamaura}
\affiliation{Research Center for Functional Materials, National Institute for Materials Science, 1-1 Namiki, Tsukuba, Ibaraki 305-0044, Japan}
\affiliation{Graduate School of Chemical Sciences and Engineering, Hokkaido University, North 10 West 8, Kita-ku, Sapporo, Hokkaido 060-0810, Japan}

\author{J. P. Hill}
\affiliation{Condensed Matter Physics and Materials Science Department and National Synchrotron Light Source-II, Brookhaven National Laboratory, Upton New York 11973, USA}

\author{J. van den Brink}
\affiliation{Institute for Theoretical Solid State Physics, IFW Dresden, D01171 Dresden, Germany}

\author{D. F. McMorrow}
\affiliation{London Centre for Nanotechnology and Department of Physics and Astronomy, University College London, Gower Street, London, WC1E 6BT, United Kingdom}

\author{A. D. Christianson}
\affiliation{Quantum Condensed Matter Division, Oak Ridge National Laboratory, Oak Ridge, Tennessee 37831, USA}
\affiliation{Department of Physics and Astronomy, University of Tennessee, Knoxville, Tennessee 37996, USA}

 

\begin{abstract}
NaOsO$_3$ hosts a rare manifestation of a metal-insulator transition driven by magnetic correlations, placing the magnetic exchange interactions in a central role. We use resonant inelastic x-ray scattering to directly probe these magnetic exchange interactions. A dispersive and strongly gapped (58 meV) excitation is observed indicating appreciable spin-orbit coupling in this 5$d^3$ system. The excitation is well described within a minimal model Hamiltonian with strong anisotropy and Heisenberg exchange ($J_1$=$J_2$=13.9 meV). The observed behavior places NaOsO$_3$ on the boundary between localized and itinerant magnetism. 
 \end{abstract}
 

\maketitle

The underlying mechanisms driving a metal-insulator transition (MIT) is an enduring focus of condensed matter physics \cite{ImadaMIT}. Recent interest has extended investigations to 5$d$-based transition metal oxides that host new paradigms of competing interactions creating novel MITs \cite{annurev-conmatphys-020911-125138}. For example spin-orbit coupling (SOC) in 5$d^5$ iridates dramatically influences the electronic ground state to allow even the presence of the reduced on-site Coulomb interaction ($U$) to drive a relativistic Mott MIT \cite{KimScience}. Conversely, in 5$d^3$ osmium-based compounds MITs occur that cannot be reconciled with the reduced $U$ in 5d systems, even when the large SOC is taken into account. These compounds therefore fall outside the Mott approximation. Of particular interest in this regard are the osmates NaOsO$_3$ and Cd$_2$Os$_2$O$_7$ which undergo a MIT that is continuous and coincident with the onset of magnetic order indicating the central role of magnetic interactions in the transition \cite{MandrusCd2Os2O7, yamaura2012, ShiNaOsO3, NaOsO3Calder}. Accessing the collective role of the competing inter and intra-ion electron-electron interactions, SOC and magnetism in driving the MIT is required to gain an understanding of the novel MITs in these osmates. 

Outside of a Mott MIT several other mechanism exist to describe the transition, including an Anderson MIT driven by disorder \cite{PhysRev.109.1492} and a Peirels MIT driven by a structural distortion in a low dimensional system \cite{Peierls}. Slater considered a route in which magnetism could drive a MIT with the central observation being that within a magnetically ordered system the potential created by an up-spin is different from that created by a down-spin \cite{slater1951}. By this definition three-dimensional magnetic ordering with oppositely aligned spins is a route to a MIT, and implicitly includes q=0 antiferromagnetic structures. NaOsO$_3$ exhibits several features consistent with Slater's general scenario \cite{ShiNaOsO3, NaOsO3Calder, du2012, jung2013, lovecchio2013, middey2014, calder2015_naoso3}. The MIT occurs concomitant with the onset of antiferromagnetic ordering ($\rm T_N=T_{MIT}=410$ K) that can create a periodic potential. Furthermore in NaOsO$_3$ the MIT is continuous and no structural symmetry change occurs. However several important questions have so far remained experimentally inaccessible hindering the development of further insight into the mechanism of this unusual MIT and prohibiting a quantitative description beyond the mean-field approach invoked by Slater for a magnetic MIT. 

Principally, since the MIT is driven by magnetic ordering the magnetic exchange interactions ($J$) are central to the creation of the MIT in NaOsO$_3$. Therefore measuring the dominant exchange pathways and interactions is required to build robust models of the MIT. Additionally the energy scales of the interactions that are required to describe the electronic behavior, such as crystal field splitting, Hund's coupling and SOC have not been accessed. In particular the nominal 5$d^3$ electronic occupancy suggests zero orbital angular momentum in the $L\!-\!S$ coupling limit and previous experimental descriptions did not require the inclusion of strong SOC. However, mounting experimental evidence in other 5$d^3$ systems indicates SOC is required to describe the magnetism \cite{PhysRevB.91.075133, TaylorSpinGap, calder2015_cd227}. To answer these questions we performed resonant inelastic x-ray scattering (RIXS) to directly probe the 5$d$ electrons of the Os ion in NaOsO$_3$.


RIXS measurements at the Os L$_3$-edge (10.87 keV) were performed on the ID20 spectrometer at the ESRF, Grenoble. A single crystal of NaOsO$_3$, space group $Pnma$, was oriented with the (H0L) plane normal to the sample surface. The scattering plane and incident photon polarization were both horizontal. The incident beam was focused to a size of 20 $\times$ 10 $\mu$m$^2$ (H$\times$V) at the sample position. Measurements were performed with low ($\Delta$E=275 meV) and high resolution ($\Delta$E = 46 meV) set-ups by switching between a Si$(311)$ channel-cut secondary monochromator and a $(664)$ four-bounce, respectively. Both set-ups used a Si$(664)$ diced spherical analyzer at 2m radius from the detector. Preliminary measurements were performed at the Advanced Photon Source (APS) on the MERIX instrument using an identical set-up to that described in Ref.~\onlinecite{calder2015_cd227}.

\begin{figure}[tb]
	\centering     
	\includegraphics[trim=0.4cm 1.5cm 0.1cm 0.1cm,clip=true, width=1.0\columnwidth]{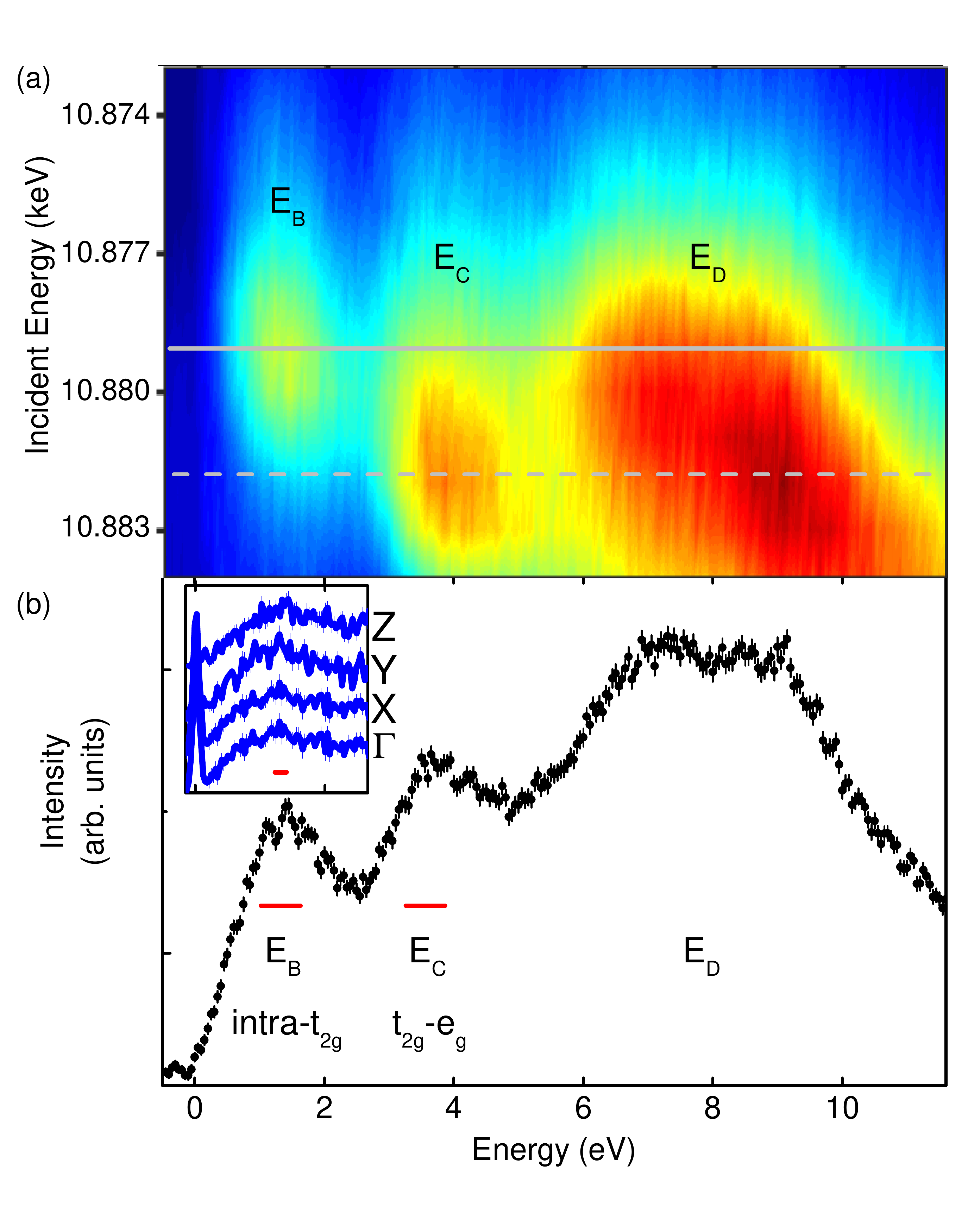}
	\caption{\label{FigRIXSmap} (a) Excitations measured in NaOsO$_3$ at various fixed incident energies through the Os L-edge using RIXS. Three inelastic peaks are observed labeled $\rm E_B$, $\rm E_C$ and $\rm E_D$. The solid (dashed) gray line indicates the incident energy that yields the maximum intensity for $\rm E_B$ ($\rm E_C$). (b) RIXS measurement at fixed incident energy of 10.88 keV. Main panel was measured with a resolution of $\Delta$E=275 meV and the inset data was collected with $\Delta$E=46 meV. The horizontal red lines indicate the FWHM experimental resolution. Measurements were performed at 300 K.}
\end{figure}

We begin by considering the results using the low resolution set-up ($\Delta$E=275 meV) before focusing on our main finding of a spin-wave excitation with high resolution measurements ($\Delta$E = 46 meV).  A RIXS map of the orbital excitations, that involve intra or inter $d$-$d$ transitions, in NaOsO$_3$ obtained by measuring the inelastic energy loss spectrum at several fixed incident energies through the Os $L_3$ resonant edge is shown in Fig.~\ref{FigRIXSmap}(a) at 300 K. Three broad inelastic features are observed, labeled $\rm E_B$, $\rm E_C$ and $\rm E_D$. On a qualitative level the excitations appear analogous to RIXS measurements on 5$d^3$-based Cd$_2$Os$_2$O$_7$ \cite{calder2015_cd227} and exhibit notable differences from measurements of 5$d^5$ based iridates \cite{PhysRevLett.108.177003}. The incident energy dependence of features $\rm E_B$ and $\rm E_C$ in the RIXS map is consistent with a nominal splitting of the 5$d$ manifold into states with $t_{\rm 2g}$ and $e_{\rm g}$ symmetry with the scattering following dipole selection rules ($\Delta \rm S=1$). $\rm E_B$ involves intra-$t_{\rm 2g}$ transitions whereas  $\rm E_C$ is due to $t_{\rm 2g}$-$e_{\rm g}$ excitations. We assign excitation $\rm E_D$ to ligand-metal charge transfer (LMCT). 

The inelastic excitation $\rm E_B$ is centered at 1.27(2) eV and corresponds to intra-$t_{\rm 2g}$ transitions.  As shown in the inset of Fig.~\ref{FigRIXSmap}(b) even when measured with the high-resolution RIXS set-up $\rm E_B$ remained as a broad single-peaked excitation, significantly broader than the instrumental resolution of 46 meV. This makes any splitting of the $t_{2g}$ manifold from SOC or a structural distortion unresolvable in NaOsO$_3$. This contrasts to the case of the iridium-based relativistic Mott insulators, where SOC strongly splits the $t_{2g}$ electronic ground state into \jeff and \jeffthree bands and results in a measured splitting of $\rm E_B$ \cite{PhysRevLett.108.177003, PhysRevLett.112.026403}.  Since $\rm E_B$ consists of intra-$t_{\rm 2g}$ transitions we follow the reasoning outlined in Ref.~\onlinecite{calder2015_cd227} to experimentally define the Hund's coupling energy ($\rm J_H$) value in NaOsO$_3$ as $\rm J_H$=1.27 eV/3.75=0.34 eV based on the center of $\rm E_B$. The 3.75 factor derives from the consideration of $\rm E_B$ as consisting of eight S=1/2 states, five of which have relative energy 3$\rm J_H$ and three having 5$\rm J_H$ yielding an average of 3.75$\rm J_H$. We note that since any underlying splitting of $\rm E_B$ is unresolved this is an approximate measurement of $\rm J_H$, however is useful for comparing with similar 5d$^3$ systems such as Cd$_2$Os$_2$O$_7$.

Excitation $\rm E_C$ is located at 3.6(1) eV and is a direct measure of the crystal field splitting. NaOsO$_3$ and Cd$_2$Os$_2$O$_7$ have a similar local OsO$_6$ octahedral environment, however the $t_{2g}$-$e_g$ splitting is 0.9 eV lower in NaOsO$_3$ \cite{calder2015_cd227}. This indicates that considerations beyond the local 5$d$ octahedral environment are crucial, which is in line with expectations of the importance of the spatially extended 5d orbitals.

\begin{figure}[tb]
	\centering     
	\includegraphics[trim=0.6cm 1.7cm 0.2cm 0.8cm,clip=true, width=0.6\columnwidth]{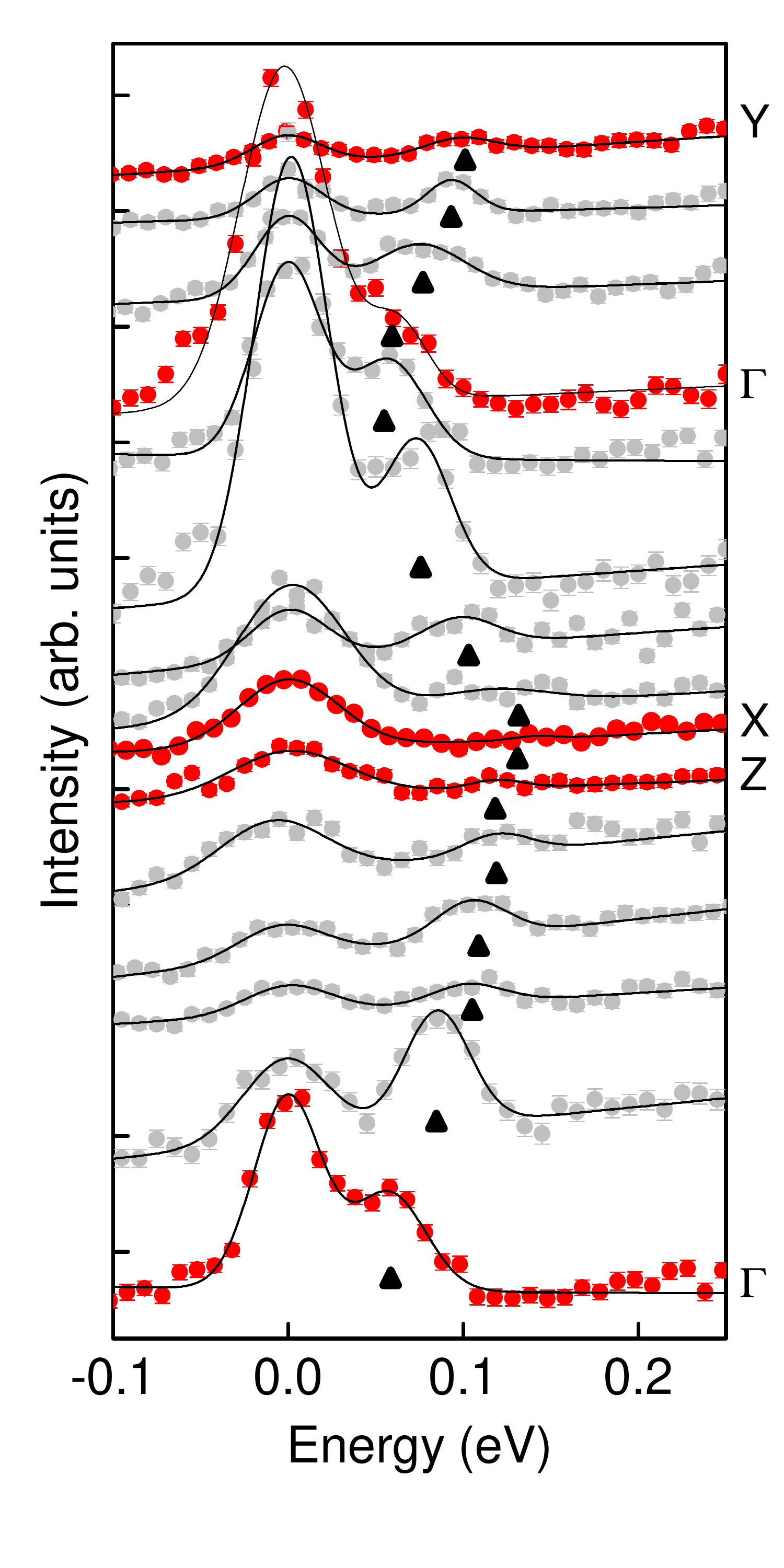}
	\caption{\label{FigMagnons} RIXS measurements of NaOsO$_3$ at 300 K along high symmetry directions in the Brillouin zone. The line shows the fit to the elastic and inelastic scattering at $\rm E_A$, with the position of  $\rm E_A$ indicated by the triangles. The data have been offset to aid comparison. The strong variation of the elastic line intensity is due to moving through the ideal 2$\theta$=$90^{\circ}$ RIXS condition where elastic scattering is suppressed.}
\end{figure}

\begin{figure}[tb]
	\centering     
	\includegraphics[trim=0.5cm 0cm 0.5cm 0cm,clip=true, width=1.0\columnwidth]{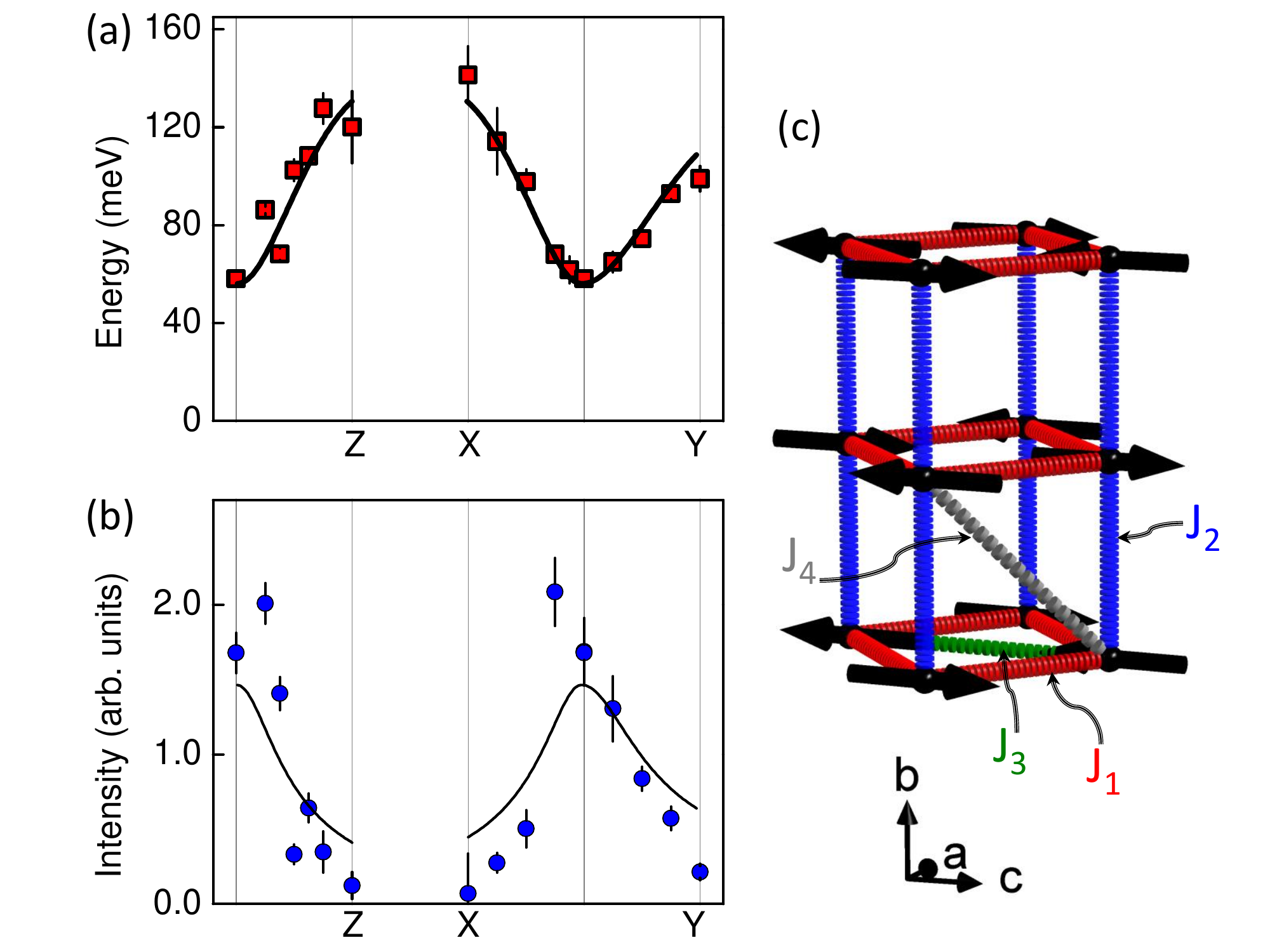}
	\caption{\label{FigDispersion} (a) Measured (squares) and calculated (line) dispersion of $\rm E_A$ along high symmetry directions. (b) Measured (circles) and calculated (line) intensity of $\rm E_A$. (c) The calculations were performed using equation (1), a minimal model Hamiltonian with the exchange interactions $J_1$=$J_2$=13.9 meV.}
\end{figure}

 The magnetic order and consequently the magnetic exchange interactions are central to the creation of the MIT in NaOsO$_3$. Therefore measuring and modeling the dominant exchange pathways is required to yield a complete picture of the MIT. The crystal size of NaOsO$_3$ is currently beyond the limits of inelastic neutron scattering, however RIXS offers a route to quantitatively probe the collective magnetic excitations in $5d$ systems \cite{PhysRevLett.108.177003}. The low energy scattering for NaOsO$_3$ using high resolution Os RIXS is shown in Fig.~\ref{FigMagnons}. Measurements along high symmetry directions shows a single resolution limited inelastic excitation ($\rm E_A$). The excitation, along with the elastic line, were fit with a Gaussian peak shape to follow the dispersion. Fig.~\ref{FigDispersion}(a) reveals $\rm E_A$ is strongly dispersive indicative of a spin-wave excitation. The bandwidth is $\sim$80 meV with maxima at the zone boundary along the Z and X direction and a reduced energy at Y. A large spin gap of 58 meV is observed at zone center ($\Gamma$) that signifies the presence of strong anisotropy in NaOsO$_3$.

To provide a quantitative description of the magnetic excitations we invoke a minimal model Hamiltonian with nearest neighbor (nn) and next nearest neighbor (nnn) exchange interactions:

\begin{equation}
\label{ham}
\mathcal{H}=J_1\sum_{nn} \mathbf{S}_i\cdot\mathbf{S}_j + J_2\sum_{nnn} \mathbf{S}_i\cdot\mathbf{S}_j + \mathbf{\Delta}
\end{equation}

A SOC induced anisotropic term is included to account for the gap. For the case of symmetric exchange anisotropy $\mathbf{\Delta}=\mathbf{\Gamma}\!\sum_{nn,nnn}\!S^z_iS^z_j$ and for single-ion anisotropy $\mathbf{\Delta}=D\sum(S_z ^2)_i$

The RIXS data for NaOsO$_3$ was modeled within a linear spin wave (LSW) approximation \cite{Spinwave} and the results checked against numerical calculations using SpinW \cite{toth2015}. The nominal spin-only value of S=3/2 was used throughout. Fitting the experimental dispersion to equation (1) produces close agreement to both the dispersion and corresponding intensity, Fig.~\ref{FigDispersion}, indicating the minimal model Hamiltonian captures the essential features and consequently provides an experimental assignment of the dominant magnetic exchange interactions and their energy. The fitting yields $J_1$=$J_2$=$13.9(5)$~meV and $\mathbf{\Gamma}$=$1.4(1)$~meV. Allowing $J_1$ and $J_2$ to vary independently did not improve the fit, reflecting the pseudo-cubic nature of the structure. Adding a third-nearest neighbor term $J_3$, or $J_3$=$J_4$ to recover the cubic limit, alters the energy at Z with respect to X, however within resolution the measured energy at X=Z. Therefore we conclude that exchange interactions $J_3$ and above have a magnitude appreciably less than $J_1$ or $J_2$ and limit the Hamiltonian to equation (1). Consequently NaOsO$_3$ is well described by dominant nearest neighbor magnetic interactions in three dimensions forming a robust G-type antiferromagnetic order in this perovskite.  Replacing exchange anisotropy with a single-ion anisotropy ($D\sum(S_z ^2)_i$) term yields the same $J_1$ and $J_2$ values and $D$=$4$ meV.

The mechanism for the spin-gap  is effectively disconnected from the magnetic exchange interactions in equation (1). However,  the presence of such a large spin-gap is significant in terms of the underlying physics in NaOsO$_3$. In particular, all possible mechanisms to open a spin-gap, single-ion anisotropy (SIA), the Dzyaloshinskii-Moriya (DM) interaction and exchange anisotropy, require SOC. We briefly consider how the anisotropies influence NaOsO$_3$. SIA arises due to a non-cubic environment and in NaOsO$_3$ the OsO$_6$ octahedra are weakly trigonally and tetragonally compressed.  However, such a large gap due to SIA does not appear consistent with the reduced spin-gaps of 15-20 meV observed in other 5$d^3$ osmates with similar distortions \cite{TaylorSpinGap}. In NaOsO$_3$ a non-zero local DM vector exists since the oxygen mediating the superexchange interaction between the two Os sites does not sit at an inversion center. Experimentally there is evidence of weak ferromagnetism \cite{ShiPRB}, however the spin-canting producing this was undetectable with neutron diffraction \cite{NaOsO3Calder}. This would suggest that the DM interaction, while present, is weak to first order. Exchange anisotropy, a pseudodipolar effect, results as a consequence of second-order SOC effects between neighboring Os ions, and hence is generally weaker than the DM interaction and SIA. However, in 5$d$ systems the extended orbitals result in enhanced collective behavior within the lattice compared to, for example, that which occurs in 3$d$ transition metal oxides. Indeed this was shown in NaOsO$_3$ with the observation of a record large spin-phonon shift at the MIT due to the extended Os orbitals \cite{calder2015_naoso3}.

The measurement of a large-spin gap indicates that SOC is required in a complete description of NaOsO$_3$. The magnitude of SOC scales with the atomic number, therefore in 5$d^3$ osmates it is comparable to 5$d^5$ iridates. However, when describing the properties of NaOsO$_3$ the role of SOC  has only been required to be included as a perturbation \cite{ShiNaOsO3, NaOsO3Calder, du2012, jung2013, lovecchio2013, middey2014, calder2015_naoso3}. Conversely for 5$d^5$ iridates SOC is necessary to describe both magnetism and the insulating state. This has created an apparent dichotomy between the effect of SOC, particularly when considering the divergent electronic ground states indicated from the RIXS spectra between 5$d^3$ and 5$d^5$. A first approximation is that the altered electronic occupancy causes an increased Hund's coupling in 5$d^3$ systems favoring a quenching of orbital momentum. However, the broad scattering observed for $E_B$ in Fig.~\ref{FigRIXSmap} indicates underlying splitting of the $t_{\rm 2g}$ orbitals, either through structural distortions or SOC or a combination of both. One consequence of an increased role of SOC in NaOsO$_3$ was considered theoretically to reduce the effective $U$ and place the system closer to the itinerant limit description of magnetism \cite{FranchiniNaOsO3}. However, the agreement of the minimal-model Hamiltonian, based on localized spins, to the RIXS spectra would suggest that the behavior of NaOsO$_3$ appreciably departs from being fully itinerant. Indeed this would be expected even within the mean-field Hartree-Fock description used by Slater to describe a magnetic MIT since at the transition local moments are necessarily formed \cite{FGebhard}. While in a pure Slater description this would be treated by self-consistent single electron theory, at least to a good approximation the behavior can be described by a Heisenberg model and places NaOsO$_3$ on the boundary between local-moment and itinerant magnetism.
 
The strongly dispersing excitation in NaOsO$_3$ contrasts with the disperionless excitation of Cd$_2$Os$_2$O$_7$ observed previously with RIXS \cite{calder2015_cd227}. The use of an Ising-like description to describe Cd$_2$Os$_2$O$_7$ and a Heisenberg model to capture the behavior for NaOsO$_3$ may indicate distinct magnetic interactions in these two closely related materials. While the behavior appears to diverge, indicating the potential for varied and exotic phenomena in related 5d$^3$ oxides, for both osmates the magnetic excitation spectra reveal an appreciable influence of SOC.  

The 5$d$ orbital and collective magnetic excitations in NaOsO$_3$ have been probed with RIXS. Well-defined spin waves were observed and described within linear spin wave theory using a minimal model Hamiltonian that captured the essential features. Nearest and next nearest neighbor Heisenberg exchange interactions of $J_1$=$J_2$=13.9 meV were found to describe the dispersion, indicating strong  three-dimensional magnetic interactions. The presence of significant anisotropy in the system was observed with the measurement of a large spin gap of 58 meV. This is a direct consequence of  intrinsically strong SOC in this 5$d$ compound, however the role of SOC on the ground state departs from 5$d^5$ iridates. In terms of the mechanism of the MIT the results support a three-dimensional magnetically driven route, consistent with the general scenario proposed by Slater. The Hamiltonian presented here provides the magnetic interaction energy scales and their pathways required to describe the MIT. Moreover, the presence of SOC and the influence this has within a system with extended orbitals and strong hybridization has to be considered as playing an important role when considering the collective interactions and the MIT.

\begin{acknowledgments}
A.D.C. thanks A. E. Taylor for useful discussions. We
thank C. Henriquet and R. Verbeni at the ESRF for the design
and manufacture of the high-temperature stage. Preliminary
measurements were performed at 9-ID-B and 30-ID (MERIX),
APS, which is a U.S. Department of Energy (DOE) Office of
Science User Facility operated for the DOE Office of Science
by Argonne National Laboratory under Contract No. DEAC02-06CH11357.
This research used resources at the High
Flux Isotope Reactor and Spallation Neutron Source, a DOE
Office of Science User Facility operated by the Oak Ridge
National Laboratory. This work in London was supported by
the UK Engineering and Physical Sciences Research Council
(EPSRC). J.G.V. would like to thank UCL and EPFL for financial
support via a UCL Impact Studentship. X.L. acknowledges
financial support from MOST (No. 2015CB921302) and CAS
(No. XDB07020200) of China. Work done at Brookhaven
National Laboratory was supported by US DOE, Division of
Materials Science, under Contract No. DE-SC00112704. K.Y.
thanks financial support from JSPS KAKENHI (15K14133
and 16H04501). This manuscript has been authored by UT-Battelle, LLC under Contract No. DE-AC05-00OR22725 with the U.S. Department of Energy. The United States Government retains and the publisher, by accepting the article for publication, acknowledges that the United States Government retains a non-exclusive, paidup, irrevocable, world-wide license to publish or reproduce the published form of this manuscript, or allow others to do so, for United States Government purposes. The Department of Energy will provide public access to these results of federally sponsored research in accordance with the DOE Public Access Plan(http://energy.gov/downloads/doepublic-access-plan).
\end{acknowledgments}


\begin{thebibliography}{25}
\bibitem [{Spi()}]{Spinwave}%
\BibitemOpen
\href@noop {} {}\bibinfo {note} {Diagonalizing the Hamiltonian in equation
	(1) within a LSW approximation gives the spin wave energy $\omega$: $\omega
	=S\sqrt{D-X-Y},\quad$ where $D = 36\Delta^2 + 48\Delta J_1 + 24\Delta J_2 +
	12J_1^2 + 2J_2^2 + 16J_1J_2$ , $X$=$4J_1^2\cos{2 \pi h} + 2J_2^2\cos{2 \pi h}
	+ 4J_1^2\cos{2 \pi l}$ , $Y$=$2J_1^2 \cos{[2 \pi (h - l)]} + 2J_1^2 \cos{[2
		\pi (h + l)]}+ 16 J_1 J_2 \cos{\pi h}\cos{\pi k} \cos{\pi l}$. In a similar
	fashion the intensity can be expressed by: $S(q, \omega)$=$S\left[\frac{D -
		(X - Y)}{D - (X + Y)}\right]\delta(\omega, \omega_q)$}\BibitemShut {NoStop}%
\end{thebibliography}

%

\end{document}